\RequirePackage{fix-cm}
\documentclass[twocolumn,epjc3]{svjour3}  
\smartqed  
\RequirePackage{graphicx}
\usepackage{amsmath}
\journalname{Eur. Phys. J. C}
\begin{document}

\title{Interpreting the $\gamma$-Ray Emission of the Cygnus X region with a Combined Cygnus OB2 and Cygnus X-3 Scenario}


\author{Caijin Xie\thanksref{addr1}
        \and
        Zhihao Ouyang\thanksref{addr1}
        \and
        Yudong Cui\thanksref{e1, addr1}
        \and
        Lili Yang\thanksref{e2, addr1, addr2, addr3}
}

\thankstext{e1}{e-mail: cuiyd@mail.sysu.edu.cn (corresponding author)}
\thankstext{e2}{e-mail: yanglli5@mail.sysu.edu.cn (corresponding author)}


\institute{School of Physics and Astronomy, Sun Yat-sen University, No.2 Daxue Rd, 519082, Zhuhai             China \label{addr1}
           \and
           Department of Physics, University of Johannesburg, PO Box 524, Auckland Park 2006, South Africa \label{addr2}
           \and
           Tianfu Cosmic Ray Research Center, Chengdu, Sichuan, People’s Republic of China\label{addr3}
}

\date{Received: date / Accepted: date}

\maketitle

\begin{abstract}
The recent detection of high-energy $\gamma$-ray emission up to PeV energies from the Cygnus X region points to the presence of a PeVatron therein. This region hosts notable candidates such as the Cygnus OB2 association, which may be associated with PeV $\gamma$-photons detected by the Large High Altitude Air Shower Observatory (LHAASO). The X-ray binary Cygnus X-3 also lies along the line of sight to the Cygnus X region, and LHAASO has observed modulated PeV $\gamma$-ray flux from its direction, indicating that its contribution is also non-negligible. Here, we attempt to interpret the $\gamma$-ray emission of the Cygnus X region using a combined scenario. Assuming a hadronic origin, we consider contributions from Cygnus OB2, Cygnus X-3, and Galactic cosmic rays (GCRs). Our results show that Cygnus OB2 alone, or its combination with the GCR component, is insufficient to account for the multiwavelength observations, whereas including the $\gamma$-ray emission from Cygnus X-3 naturally explains these observations. This suggests that the $\gamma$-ray emission from the Cygnus X region has a more complex origin, necessitating a joint consideration of Cygnus OB2 and Cygnus X-3. The expected neutrino flux from hadronic interactions extends into the GeV band and can help reveal the radiation mechanism of Cygnus X-3, underscoring the need for high-sensitivity GeV neutrino observatories. 

\end{abstract}

\section{Introduction}\label{intro}
Cygnus X, located at a distance of about 1.46 kpc, is one of the most active star-forming regions in the Milky Way, harboring several Wolf--Rayet stars and hundreds of O-type stars grouped within massive OB associations. The powerful wind termination shock (WTS) can boost particles to TeV--PeV energies. 
Along with other candidate cosmic-ray accelerators such as massive stars and supernova remnants (SNRs), these conditions establish Cygnus X as a promising Galactic particle accelerator. The large-scale molecular and atomic clouds in its vicinity supply abundant target gas for proton–proton ($pp$) interactions, making the region highly favorable for the production of 
$\gamma$-rays and neutrinos through hadronic processes.
Recently, the Large High Altitude Air Shower Observatory (LHAASO) reported high-energy $\gamma$-ray emission extending to PeV energies from the Cygnus X region \cite{1}. The detected $\gamma$-photons are distributed within a region of $\sim6^{\circ}$ radius centered on Cygnus X and exhibit a $1/r$ type radial distribution. IceCube Collaboration has also reported an excess of neutrino events from the inner Galactic plane ($|l|<60^{\circ}$, $|b|<8^{\circ}$) \cite{49,50}, particularly from the Cygnus X region \cite{42}. These observations strongly support the presence of a PeVatron at the center of Cygnus X that produces $\gamma$-rays and neutrinos via hadronic interactions.

One of the most significant PeVatron candidates in Cygnus X is the Cygnus OB2 association \cite{1,8}. As the nearest and most massive young star cluster in the Milky Way, it contains 169 confirmed OB stars, including 52 O-type and 3 Wolf–Rayet stars \cite{2}. The large population of massive stars drives powerful stellar winds whose shocks, formed through interactions with the interstellar medium (ISM) or between neighboring wind shells, provide favorable conditions for particle acceleration.
In addition to WTS, collective winds and supernova explosions within the cluster are also viable accelerators in Cygnus OB2. Surveys of molecular and atomic gas, including the CfA 1.2 m CO survey \cite{3} and the HI4PI Survey \cite{4}, reveal abundant clouds at distances consistent with Cygnus OB2, providing ample target material for hadronic interactions. High-energy protons accelerated by the cluster can diffuse into these clouds and produce $\gamma$-rays via inelastic $pp$ collisions. The resulting emission from the $\sim1^{\circ}$ region surrounding Cygnus OB2, known as the “Cygnus Cocoon”, was first detected by \textit{Fermi}-LAT \cite{5}, with subsequent observations by ARGO \cite{6}, HAWC \cite{7}, and LHAASO \cite{8} extending the measurements into the TeV–PeV range. These multi-wavelength observations provide compelling evidence for ongoing particle acceleration in Cygnus OB2.

Along the line of sight to the Cygnus X region lies the X-ray binary Cygnus X-3, at a distance of approximately 9.0 kpc. This system comprises a compact object, likely a black hole or a neutron star, hosting a highly relativistic jet \cite{9}, together with a companion Wolf--Rayet star. Its orbital period is only 4.8 hours, and modulation has been observed in the radio, infrared, X-ray, and GeV $\gamma$-ray bands \cite{10,11,12}. In addition to this short period, the source is characterized by an exceptionally high X-ray luminosity, which would exceed $5.5\times10^{39}$ erg s$^{-1}$ \cite{13}. Such intense X-ray emission implies a strong jet kinetic power sufficient to accelerate particles to PeV energies. The companion Wolf--Rayet star can supply cold protons for $pp$ interactions with nonthermal protons in the jet, while the intense radiation fields around the central compact object can drive photohadronic ($p\gamma$) interactions with these high-energy protons. This scenario is supported by the recent LHAASO detection of modulated PeV $\gamma$-rays from the direction of Cygnus X-3 \cite{14}, making this X-ray binary a promising PeVatron candidate \cite{15,16,33}. 

In this work, we propose a combined scenario to interpret the $\gamma$-ray emission from the Cygnus X region. We first consider protons accelerated by the WTS of Cygnus OB2 that subsequently undergo $pp$ interactions with nearby molecular and atomic clouds. Even with Bohm diffusion adopted, this component is significantly insufficient to account for the PeV $\gamma$-ray flux of Cygnus X. Since Cygnus OB2 resides in a gas-rich environment, Galactic cosmic rays (GCRs) can also interact with this gas through $pp$ collisions and produce $\gamma$-rays. We adopt gas densities consistent with those derived from the CfA 1.2 m CO survey and the HI4PI survey \cite{1}; however, the combined contribution from Cygnus OB2 and GCRs still remains deficient at PeV energies. As LHAASO detected a modulation effect in the $\gamma$-ray observations near Cygnus X-3, which substantially increases the probability that these $\gamma$-rays originate from Cygnus X-3, the contribution of Cygnus X-3 is also non-negligible. We model the $\gamma$-ray emission from the core region of Cygnus X-3, which is dominated by the relativistic jet, via both $pp$ and $p\gamma$ interactions under constraints from the LHAASO observation \cite{14}. Moreover, the nonthermal particles in the jet are assumed to be able to escape from the jet, diffuse into the ISM, and generate diffuse $\gamma$-rays. With both core and diffuse emission considered, the shortfall at PeV energies is naturally addressed, and the multiwavelength spectrum of the Cygnus X region is well explained with conventional parameters. This result favors Cygnus X-3 as the PeVatron of Cygnus X. The accompanying neutrino flux from hadronic interactions is also calculated, and its spectrum extends into the GeV band. Although the $\gamma$-ray emission from Cygnus X-3 in the 10 TeV to 100 TeV band can be explained by either $pp$ or $p\gamma_{\rm{X}}$ interactions \cite{14} and these two channels are degenerate in the $\gamma$-ray spectrum, their neutrino signals differ in the GeV band. This implies that GeV neutrino observations can help identify the physical origin of the $\gamma$-rays from Cygnus X-3, further highlighting the need for high-sensitivity GeV neutrino observatories. 

\section{Particle acceleration at the WTS of Cygnus OB}\label{part}
The supersonic cluster wind of Cygnus OB2 impacts the surrounding ISM and inflates a superbubble (SB). Within this SB, winds from individual stars can interact with the ISM or with winds from other stars, generating a termination shock (TS). The particles are accelerated at the TS and then diffuse throughout the SB. Upstream of the TS, the cluster wind is advected at a uniform speed $u=M_{A}V_{A}$, where $M_{A}$ represents the Mach number and $V_{A}$ denotes the Alfv\'{e}n speed. Downstream of the TS, the shocked cluster wind is advected outward with a radially dependent velocity. The TS radius $R_{s}$ and SB radius $R_{b}$ are assumed to be 15 pc and 96 pc \cite{20}, respectively. Since LHAASO has detected $\gamma$-rays extending to an angular distance of $\sim6^{\circ}$ from the center of Cygnus X \cite{1}, we assume that gas is distributed within a radius $R_{c}=155$ pc, corresponding to $6^{\circ}$ at a distance of 1.46 kpc. Assuming the system to be spherically symmetric, the overall particle acceleration and diffusion process is conventionally described by a transport equation \cite{20,17,18,19}
\begin{align}\label{eq-1}
    \frac{\partial N(r,E,t)}{\partial t}=\;&\frac{1}{r^{2}}\frac{\partial}{\partial r}\left[r^{2}D(E)\frac{\partial N(r,E,t)}{\partial r}\right]\nonumber\\&-\frac{1}{r^{2}}\frac{\partial}{\partial r}[r^{2}\tilde{u}(r)N(r,E,t)]\nonumber\\&+\frac{\partial}{\partial E}[\dot{E}N(r,E,t)]+Q(r,E,t).
\end{align}

Here, $N(r,E,t)$ is the particle number per unit energy per unit volume. The term $D(E)$ stands for the spatial diffusion coefficient. Inside the SB, it can be written as
\begin{equation}\label{eq-2}
    D(E)=\frac{1}{3}cL_{c}^{1-\delta}r_{L}^{\delta}(E), 
\end{equation}
where $L_{c}=1$ pc is the coherence scale of the perturbations \cite{20} and $r_{L}(E)\approx3.5\times10^{17}(E/1\,\mathrm{PeV})(1\,\mathrm{\mu G}/B)$ cm is the Larmor radius. The index $\delta$ depends on the diffusion regime: for Kolmogorov diffusion, $\delta=1/3$; for Kraichnan diffusion, $\delta=1/2$; and for Bohm diffusion, $\delta=1$. Outside the SB, we take $D(E)=4.0\times10^{26}\allowbreak(E/1\,\mathrm{GeV})^{1/3}$ cm$^{2}$ s$^{-1}$, approximately two orders of magnitude lower than the typical Galactic value and supported by the evidence for suppressed diffusion around the young massive star cluster RCW 38 \cite{23}. The term $\tilde{u}(r)$ denotes the effective advection velocity given by 
\begin{equation}\label{eq-3}
	\tilde{u}(r)=\begin{cases}
    \tilde{u}_{1}&R_{\rm{sc}}<r\leq R_{s} \\
    \tilde{u}_{2}\left(\frac{r}{R_{s}}\right)^{-2}&R_{s}<r\leq R_{c} 
\end{cases},
\end{equation}
where $R_{\rm{sc}}=6.4$ pc represents the radius of the central star cluster of Cygnus OB2 \cite{55}. The terms $\tilde{u}_{1}=u$ and $\tilde{u}_{2}=u/4+\eta V_{A}$ are the effective advection velocities immediately upstream and downstream of the TS. The coefficient $\eta$ quantifies the effect of a net motion of the scattering centers relative to the Alfv\'{e}n speed scale. The energy loss term $\dot{E}=\dot{E}_{\rm{ad}}+\dot{E}_{\rm{pp}}$ comprises two channels for particle energy loss\textemdash adiabatic expansion and $pp$ collisions\textemdash which can be expressed as follows \cite{20}
\begin{equation}\label{eq-4}
    \dot{E}_{\rm{ad}}=\frac{1}{3}[\nabla\cdot\tilde{u}(r)]E.
\end{equation}
\begin{equation}\label{eq-5}
    \dot{E}_{\rm{pp}}=K_{\rm{pp}}(1+0.055L+0.0073L^{2})n_{g}E.
\end{equation}
Here, $K_{\rm{pp}}=6.63\times10^{-16}$ cm$^{3}$ s$^{-1}$ and $L=\ln(E/1\,\mathrm{TeV})$. From the CfA 1.2 m CO survey and the HI4PI survey, the molecular and atomic clouds in the vicinity of Cygnus OB2 have densities of 17 cm$^{-3}$ and 4.5 cm$^{-3}$ \cite{1}, respectively. Since the molecular cloud is non-uniformly distributed and relatively concentrated within a region of $\sim3^{\circ}$ radius, we reduce its density by a factor of 4 and adopt a total gas density of $n_{g}=17/4+4.5=8.75$ cm$^{-3}$, consistent with the best-fit value from a recent study \cite{32}. The injection function
\begin{equation}\label{eq-6}
    Q(r,E,t)=4\pi R_{s}^{2}\eta_{\rm{inj}}n_{g}u\delta(E-E_{\rm{inj}})\delta(r-R_{s}),
\end{equation}
describes particle injection at the TS, with $\eta_{\rm{inj}}$ and $E_{\rm{inj}}$ reflecting the injection fraction and injection energy, respectively.

One of the key parameters governing the evolution of the proton spectrum is the diffusion coefficient $D(E)$. As shown in Eq.~(\ref{eq-2}), $D(E)$ within the SB is determined by the magnetic field $B=\sqrt{4\pi\eta_{B}\rho\tilde{u}^{2}}$, where $\eta_{B}$ represents the fraction of the wind kinetic energy channeled into turbulent magnetic energy and $\rho$ denotes the density of the wind. Here, we adopt $\eta_{B}=0.05$, comparable to the conversion efficiency inferred from the intracluster medium \cite{21}. This value yields a Mach number $M_{A}=\eta_{B}^{-1/2}\approx4.5$, indicating a strong shock that is sufficient for particle acceleration. For a strong shock, we assume that the magnetic field strength immediately downstream of the TS is enhanced by a factor of $\sqrt{11}$ relative to that immediately upstream \cite{17}. Since $\tilde{u}$ depends on $r$ downstream of the TS but is constant upstream, the diffusion coefficient should be treated within a three-region prescription: upstream and outside the SB, it remains constant; downstream, it increases with radius as $D(E)\propto r^{2\delta}$. Another key parameter influencing the evolution of the proton spectrum is $\eta$, which reflects the net motion of the scattering centers. With increasing $\eta$, a higher net motion speed would be superimposed on the downstream advection velocity, reducing the shock compression ratio between the upstream and downstream regions and thereby lowering the shock acceleration efficiency and softening the spectrum.

Under the steady-state assumption, the left-hand side of Eq.~(\ref{eq-1}) vanishes. With the method of \cite{18}, Eq.~(\ref{eq-1}) can be transformed into a set of tridiagonal equations and subsequently solved using the conventional chasing method. To simulate proton acceleration and diffusion at the WTS of Cygnus OB2, we divide the solution domain into the upstream and downstream regions of the TS and solve Eq.~(\ref{eq-1}) separately in each. For the upstream region, we adopt the vanishing flux boundary condition at the inner boundary $r=R_{\rm{sc}}$, while for the downstream region, the free-escape boundary condition is imposed at the outer boundary $r=R_{c}$. At $r=R_{s}$, a semi-analytical boundary condition is applied to both regions. Further details on solving the transport equation are provided in \ref{appA}. Using the aforementioned parameters, we compute the steady-state proton spectra across the entire SB of Cygnus OB2 and employ the \texttt{GAMERA} toolkit \cite{22} to obtain the corresponding $\gamma$-ray spectra, which are then summed to derive the integrated spectrum of Cygnus X. The details of the $\gamma$-ray spectrum calculation are given in \ref{appB}. 

\begin{figure}[!t]
    \centering
    \includegraphics[width=0.5\textwidth]{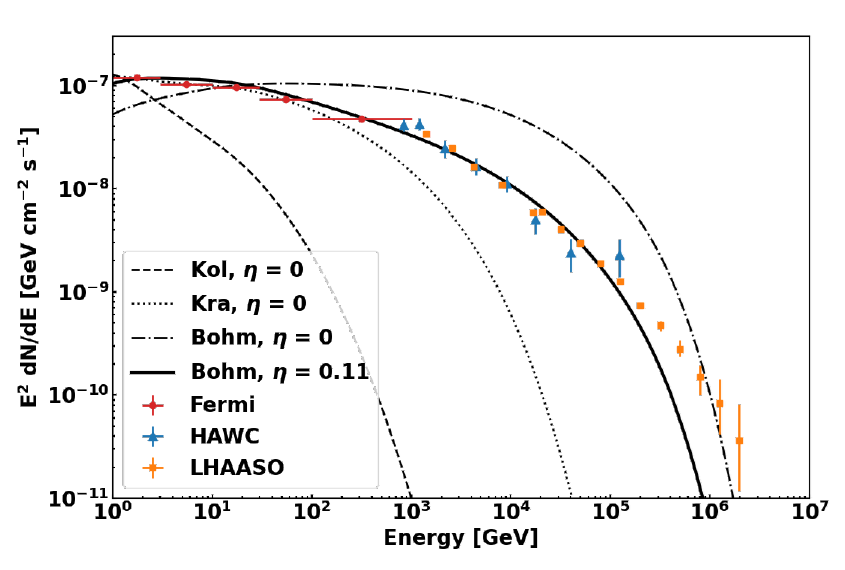}
    \caption{\label{fig:1}Modeled total $\gamma$-ray emission from Cygnus X, assuming that the protons accelerated by the WTS of Cygnus OB2 are the sole source, compared with multiwavelength observations. The dashed and dotted lines show the results for Kolmogorov and Kraichnan diffusion with $\eta=0$, respectively, while the dash-dotted and solid lines correspond to Bohm diffusion with $\eta=0$ and $\eta=0.11$. The red circles, blue triangles, and orange squares, all with error bars, indicate the latest GeV data derived from the \textit{Fermi} 4FGL catalog \cite{20}, TeV measurements by HAWC \cite{7}, and LHAASO detections up to PeV energies \cite{1}, respectively.}
\end{figure}

Figure~\ref{fig:1} presents the modeled $\gamma$-ray emission from the entire Cygnus X region when only the WTS of Cygnus OB2 is considered. The results for Kolmogorov, Kraichnan, and Bohm diffusion with different adopted values of $\eta$ are compared with multiwavelength observations from \textit{Fermi}-LAT \cite{20}, HAWC \cite{7}, and LHAASO \cite{1}. As shown, under Kolmogorov and Kraichnan diffusion, high-energy protons undergo highly efficient diffusive escape from the Cygnus OB2 SB, yielding a cutoff energy far below that required to account for the PeV $\gamma$-ray flux. For Bohm diffusion with $\eta=0$, the spectrum is excessively hard, and the expected TeV $\gamma$-rays would significantly exceed the HAWC and LHAASO data. With a sufficiently large $\eta=0.11$, the data below 100 TeV can be well reproduced, whereas the higher-energy portion remains difficult to account for. These findings are consistent with previous work \cite{20} and suggest that protons accelerated by the WTS of Cygnus OB2 alone cannot be the sole contributor to the total $\gamma$-ray emission from Cygnus X; under the WTS acceleration scenario, Cygnus OB2 is also unlikely to be the PeVatron of Cygnus X. 

\section{GCR interact with the gas around the Cygnus OB SB}\label{gcr}

\begin{figure}[!t]
    \centering
    \includegraphics[width=0.5\textwidth]{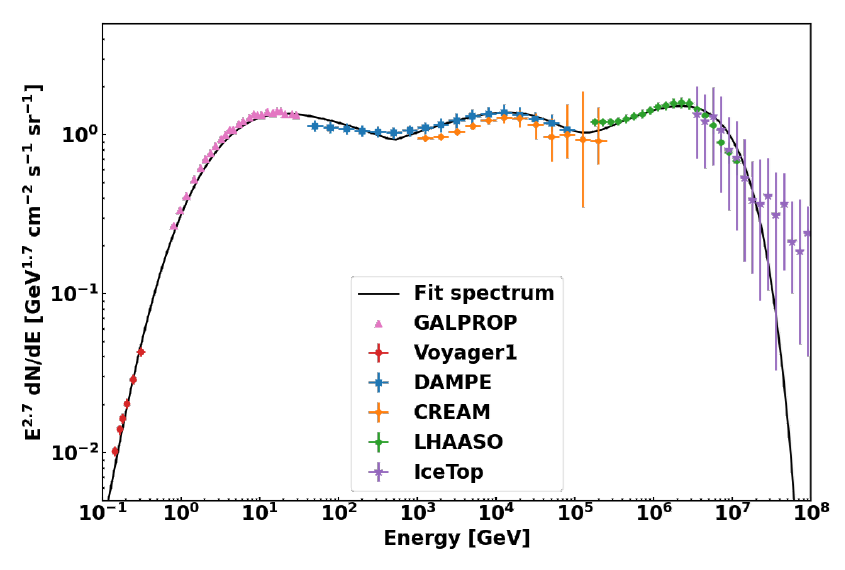}
    \caption{\label{fig:2}GCR proton spectrum measured locally. The black solid line is the fit spectrum used for GCR proton propagation simulation. The pink triangles represent the GALPROP computation result \cite{24}. The red circles, blue squares, orange diamonds, green hexagons, and purple stars, all with error bars, denote the measurement results from Voyager 1 \cite{25}, DAMPE \cite{26}, CREAM \cite{27}, LHAASO \cite{28}, and IceTop \cite{29}, respectively}
\end{figure}

In addition to protons accelerated by the WTS of Cygnus OB2, GCR protons can also produce $\gamma$-rays through $pp$ interactions\textemdash mainly with the clouds outside the Cygnus OB2 SB, while a fraction above 10 TeV penetrate into the SB and collide with the local gas\textemdash and contribute to the $\gamma$-ray emission from the Cygnus X region. The propagation of GCR protons is also governed by Eq.~(\ref{eq-1}), with the boundary conditions slightly modified. The vanishing flux boundary condition is retained at the inner boundary $r=R_{\rm{sc}}$; at the outer boundary $r=R_{c}$, we assume that the GCR proton spectrum matches the locally measured one. The local GCR proton spectrum adopted in this work is shown in Fig.~\ref{fig:2}. This spectrum is obtained by fitting the results from GALPROP and measurements from Voyager 1 \cite{25}, the Dark Matter Particle Explorer (DAMPE) \cite{26}, the balloon-borne Cosmic Ray Energetics And Mass (CREAM) experiment \cite{27}, LHAASO \cite{28}, and IceTop \cite{29}. The fit employs the template of \cite{30,31}. For $E<500$ GeV,
\begin{align}\label{eq-7}
J_{\rm{CR}}(E)=\;&a_{0}\left\{1+\exp\left[-\frac{\ln(E/E_{b})-a_{1}}{a_{2}}\right]\right\}^{a_{3}}\nonumber\\
&\times\left(\frac{E}{E_{b}}\right)^{a_{4}}
\left[1+\left(\frac{E}{a_{5}}\right)^{a_{6}}\right]^{a_{7}},
\end{align}
and for $E\geq500$ GeV,
\begin{align}\label{eq-8}
    J_{\rm{CR}}(E)=\;&A_{1}\left(\frac{E}{1\,\mathrm{TeV}}\right)^{-B_{1}}\exp\left(-\frac{E}{C_{1}}\right)\nonumber\\&+A_{2}\left(\frac{E}{1\,\mathrm{TeV}}\right)^{-B_{2}}\exp\left(-\frac{E}{C_{2}}\right).
\end{align}
Here, $E_{b}=0.45$ GeV, $a_{0}=31.83866$ GeV$^{-1}$ cm$^{-2}$ s$^{-1}$ sr$^{-1}$, $a_{1}=2.3$, $a_{2}=0.9073$, $a_{3}=-0.9976$, $a_{4}=-0.9$, $a_{5}=0.41575$ GeV, $a_{6}=1.437$, $a_{7}=-1.374$, $A_{1}=6.1\times10^{-9}$ GeV$^{-1}$ cm$^{-2}$ s$^{-1}$ sr$^{-1}$, $B_{1}=2.55$, $C_{1}=4.5\times10^{4}$ GeV, $A_{2}=2.25\times10^{-9}$ GeV$^{-1}$ cm$^{-2}$ s$^{-1}$ sr$^{-1}$, $B_{2}=2.45$, and $C_{2}=9.0\times10^{6}$ GeV are fitting parameters.

\begin{figure}[!t]
    \centering
    \includegraphics[width=0.5\textwidth]{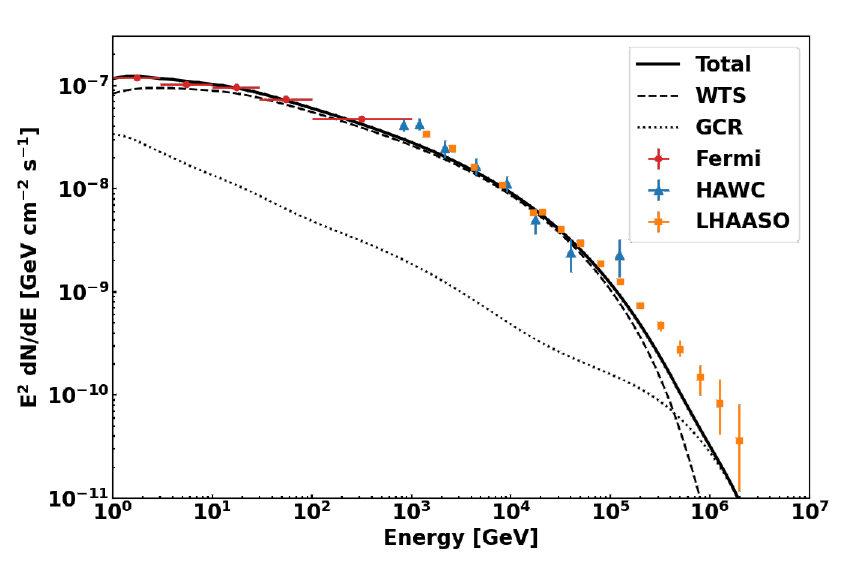}
    \caption{\label{fig:3}Modeled total $\gamma$-ray emission from Cygnus X, assuming that the protons accelerated by the WTS of Cygnus OB2 and the GCR protons are the combined source, compared with multiwavelength observations. The dashed and dotted lines show the $\gamma$-ray fluxes from WTS-accelerated protons and GCR protons, respectively, whereas the solid line indicates their sum. The observational data points are the same as those in Fig.~\ref{fig:1}.}
\end{figure}

We calculate the contribution of GCR protons to the $\gamma$-ray emission from the Cygnus X region. As shown in Fig.~\ref{fig:3}, adding this component increases the flux above 100 TeV, yet a deficit persists at PeV energies, even allowing for the large error bar of the LHAASO data in this range. In the calculation of the $\gamma$-ray flux from WTS-accelerated protons, we reduce the injection fraction $\eta_{\rm{inj}}$ by a factor of 1.25 to ensure that the low-energy part of the spectrum aligns with the \textit{Fermi} data, while keeping the other parameters identical to those used in Section~\ref{part}. The shortfall in the PeV band could be overcome by increasing the gas density $n_{g}$. However, \cite{32} reports a best-fit value of $n_{g}=6.58$ cm$^{-3}$, slightly lower than that adopted in this work. In addition, the atomic cloud density derived from the Canadian Galactic Plane Survey (CGPS) \cite{52}, the Wilkinson Microwave Anisotropy Probe (WMAP) \cite{53}, and Planck \cite{54} is expected to decrease rapidly beyond an angular distance of $\sim5^{\circ}$, where the GCR protons have the highest density. Thus, the GCR contribution obtained here can be interpreted as an upper limit, rather than as a value that can be further increased. For this reason, we favor supplementing the high-energy portion of the multiwavelength spectrum with additional sources, especially Cygnus X-3. 

\section{The contribution from Cygnus X-3}\label{contri}

\subsection{Core emission}\label{core}
The recent work \cite{14} reports the detection of $\gamma$-ray flux from the direction of Cygnus X-3, with a spectrum extending to energies of several PeV, which could originate from Cygnus X-3. Moreover, a modulation effect has been observed in the $\gamma$-ray flux above 0.1 PeV, which can be explained by the orbital motion of this binary system \cite{14,33}. This further strengthens the association between the $\gamma$-ray emission and Cygnus X-3. Following \cite{14}, we assume that the observed $\gamma$-ray flux comes from the core region of Cygnus X-3 and model the $\gamma$-ray spectrum under two scenarios. In the first scenario, nonthermal protons in the relativistic jet of the central object undergo $pp$ interactions with other jet protons and/or with the stellar wind of the Wolf--Rayet star, together with $p\gamma$ interactions between the jet protons and the UV photons from the Wolf--Rayet companion. In the second scenario, the jet protons undergo $p\gamma$ interactions with the X-ray photons in the core region, combined with $p\gamma$ interactions with the UV radiation field from the Wolf--Rayet donor.

We assume a power-law spectrum with an exponential cutoff for the nonthermal protons in the jet, written as
\begin{equation}\label{eq-9}
    n(E)=n_{0}\left(\frac{E}{1\,\mathrm{GeV}}\right)^{-\Gamma}\exp\left(-\frac{E}{E_{c}}\right),
\end{equation}
where $E_{c}$ is the cutoff energy, set to 50 PeV \cite{14}, and $\Gamma$ is the spectral index. The normalization $n_{0}$ is determined by the total energy budget $W_{p}$ of protons above 1 TeV in the relativistic jet and can be derived by 
\begin{align}\label{eq-10}
    n_{0}=\;&\frac{3W_{p}}{4\pi R_{e}^{3}}\nonumber\\&\times\left[\int_{1\,\mathrm{TeV}}^{\infty}\mathrm{d}E'E'\left(\frac{E'}{1\,\mathrm{GeV}}\right)^{-\Gamma}\exp\left(-\frac{E'}{E_{c}}\right)\right]^{-1},
\end{align}
where $R_{e}=1.0\times10^{11}$ cm represents the scale height of the emission site. 

\begin{figure*}[!t]
    \centering
    \includegraphics[width=1.0\textwidth]{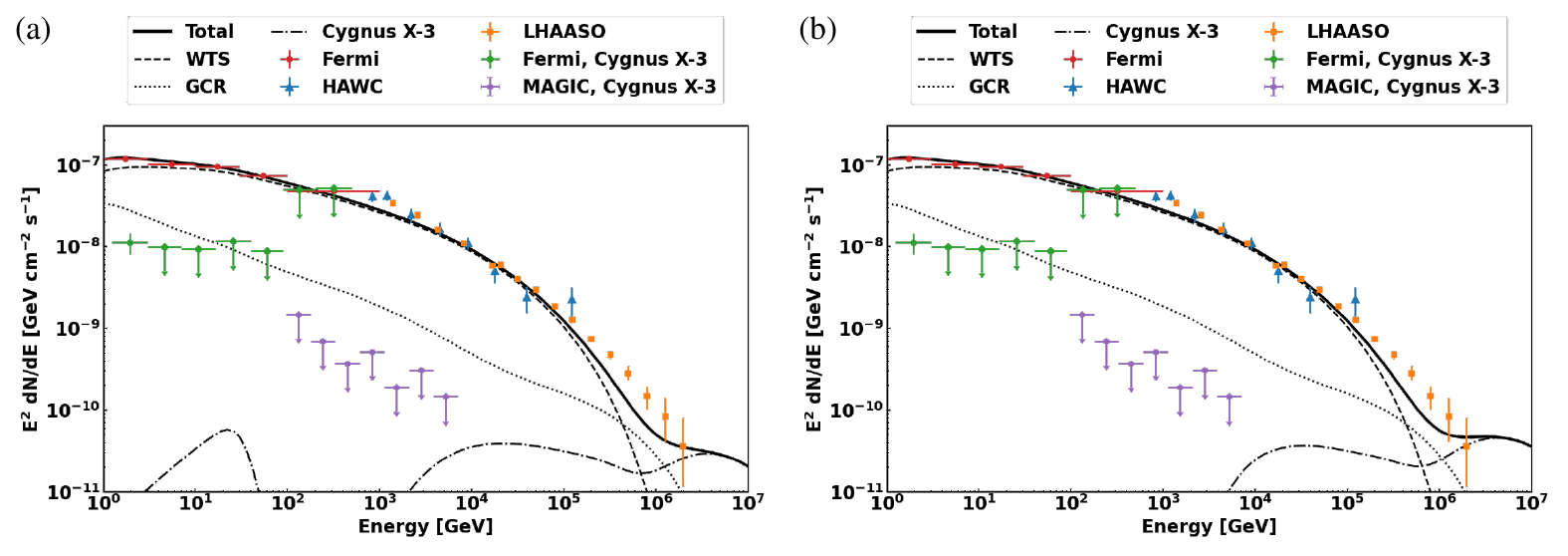}
    \caption{\label{fig:4}Modeled total $\gamma$-ray emission from Cygnus X, combining contributions from Cygnus OB2, GCR, and the core emission of Cygnus X-3, compared with multiwavelength observations. The Cygnus X-3 core emission is modeled under both the $pp+p\gamma_{\rm{UV}}$ and $p\gamma_{\rm{X}}+p\gamma_{\rm{UV}}$ scenarios, with the results shown in Panels (a) and (b), respectively. The dashed, dotted, and dash-dotted lines show the $\gamma$-ray fluxes from WTS-accelerated protons, GCR protons, and the Cygnus X-3 core emission, whereas the solid line indicates their sum. The green diamonds and purple hexagons are the observational results for Cygnus X-3 from \textit{Fermi}-LAT and MAGIC \cite{37}.}
\end{figure*}

In the first scenario, which considers $pp$ and $p\gamma_{\rm{UV}}$ interactions, the energy budget $W_{p}$ is adopted as $6.6\times10^{39}$ erg and the spectral index $\Gamma$ is set to 2.4 \cite{14}. The target proton density is assumed to be $2.0\times10^{10}$ cm$^{-3}$. The UV spectrum of the Wolf--Rayet star is taken to be a pure thermal spectrum given by
\begin{equation}\label{eq-11}
    F(E)=\frac{8\pi E^{2}R_{\rm{star}}^{2}}{h^{3}c^{3}R_{W}(R_{W}-R_{\rm{star}})}\frac{1}{\exp[E/(k_{B}T)]-1}.
\end{equation}
The temperature $T$ is set to $3.75\times10^{4}$ K. The radius $R_{\rm{star}}$ of the Wolf--Rayet star and its distance $R_{W}$ to the central object are taken as $1.0\times10^{11}$ cm and $2.65\times10^{11}$ cm \cite{14}, respectively. In the second scenario, which considers $p\gamma_{\rm{X}}$ and $p\gamma_{\rm{UV}}$ interactions, $W_{p}$ is adopted as $4.8\times10^{39}$ erg and $\Gamma$ is chosen to be 2.3 \cite{14}. The target X-ray spectrum is derived from the Monitor of All-sky X-ray Image (MAXI) \cite{34} observations between January 1, 2020, and January 1, 2026. The UV spectrum retains the form given in Eq.~(\ref{eq-11}). 

Due to the complex radiation environment near the core of Cygnus X-3, strong $\gamma\gamma$ absorption \cite{35} can occur in the core and reduce the $\gamma$-ray flux at Earth. We compute the $\gamma\gamma$ absorption optical depth $\tau_{\gamma\gamma}(E)$ using the following equation 
\begin{equation}\label{eq-12}
    \tau_{\gamma\gamma}(E)=d_{\gamma\gamma}\int_{m_{e}c^{2}}^{\infty}\mathrm{d}E_{t}\sigma_{\gamma\gamma}(E,E_{t})F(E_{t}),
\end{equation}
where $d_{\gamma\gamma}$ is the absorption length, $m_{e}$ is the electron mass, and $E_{t}$ is the energy of the target photons for $\gamma\gamma$ absorption. The $\gamma\gamma$ absorption cross-section $\sigma_{\gamma\gamma}(E,E_{t})$ can be expressed as \cite{35}
\begin{align}\label{eq-13}
    \sigma_{\gamma\gamma}(E, E_{t})=\;&\frac{1}{2}\pi r_{e}^{2}[1-\beta^{2}(E, E_{t})]\nonumber\\&\times\left\{[3-\beta^{4}(E, E_{t})]\ln\frac{1+\beta(E, E_{t})}{1-\beta(E, E_{t})}\right.\nonumber\\&\left.-2\beta(E, E_{t})[2-\beta^{2}(E, E_{t})]\right\}.
\end{align}
Here, $r_{e}$ represents the classical electron radius. The term $\beta(E, E_{t})$ denotes the velocity of the electron or positron produced after absorption in the observer frame, calculated by
\begin{equation}\label{eq-14}
    \beta(E, E_{t})=\sqrt{1-\frac{m_{e}^{2}c^{4}}{EE_{t}}}.
\end{equation}
The primary radiation fields acting as targets for $\gamma\gamma$ absorption are the IR-to-mm and X-ray emission from the central object and the UV emission from the Wolf--Rayet star. According to our calculations, compared to the optical depth due to UV absorption, the contributions from the other channels can be neglected. Therefore, following \cite{14}, we use the optical depth of the UV absorption to describe $\gamma\gamma$ absorption in the core region.

The $\gamma$-ray spectrum consistent with the high-state observational data reported in \cite{14} can be reproduced by both scenarios using the aforementioned parameters. For $p\gamma$ interactions, we adopt the simplified model B of \cite{36}. Taking orbital modulation into account, we assume that the average $\gamma$-ray flux of Cygnus X-3 is half of its high-state value. The contribution from the Cygnus X-3 core emission is combined with those from Cygnus OB2 and GCR. Figure~\ref{fig:4} presents the results. Our expected $\gamma$-ray flux from Cygnus X-3 is consistent with the upper limits given by \textit{Fermi}-LAT and Major Atmospheric Gamma-ray Imaging Cherenkov (MAGIC) \cite{37}. The total $\gamma$-ray flux of Cygnus X above PeV energies can be explained by the core emission of Cygnus X-3, whereas a shortfall remains at several hundred TeV, since the core emission from Cygnus X-3 at these energies is relatively weak. This suggests that including only the Cygnus X-3 core emission is still insufficient.

\subsection{Diffuse emission}\label{diff}

\begin{figure*}[!t]
    \centering
    \includegraphics[width=1.0\textwidth]{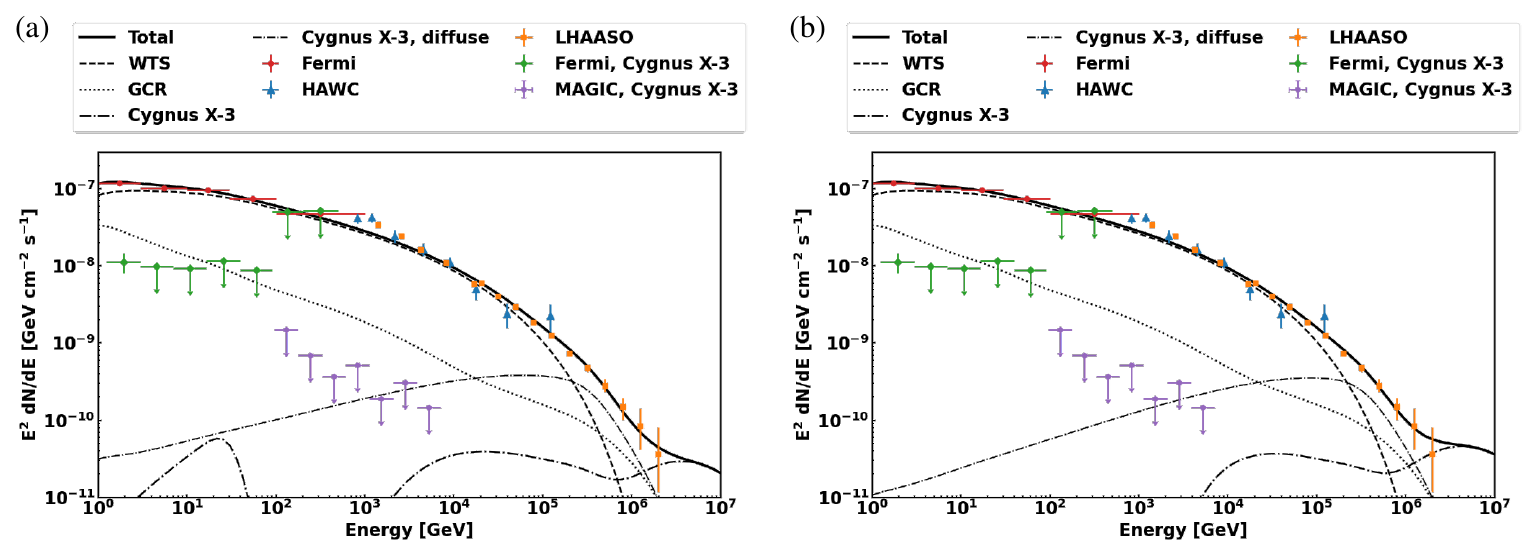}
    \caption{\label{fig:5}Modeled total $\gamma$-ray emission from Cygnus X, combining contributions from Cygnus OB2, GCR, and the core and diffuse emissions of Cygnus X-3, compared with multiwavelength observations. The Cygnus X-3 core emission and the corresponding diffuse emission are modeled under both the $pp+p\gamma_{\rm{UV}}$ and $p\gamma_{\rm{X}}+p\gamma_{\rm{UV}}$ scenarios, with the results shown in Panels (a) and (b), respectively. The dashed, dotted, thick dash-dotted, and thin dash-dotted lines show the $\gamma$-ray fluxes from WTS-accelerated protons, GCR protons, the Cygnus X-3 core emission, and the Cygnus X-3 diffuse emission, whereas the solid line indicates their sum. The observational data points are the same as those in Fig.~\ref{fig:4}.}
\end{figure*}

In addition to the core emission, Cygnus X-3 can contribute to the $\gamma$-ray emission from the Cygnus X region through another channel. Nonthermal protons in the jet can eventually escape into the ISM around Cygnus X-3, where they undergo $pp$ collisions with the ambient gas and generate a $\gamma$-ray halo \cite{16,38,39}. Here, we assume that only a fraction $\eta_{\rm{esc}}$ of the nonthermal protons is able to escape from the jet before cooling, and that the spectrum of these escaping protons retains a form similar to that in Eq.~(\ref{eq-9}). The injection and propagation of these protons in the ISM are also governed by Eq.~(\ref{eq-1}). Assuming that the protons escape into the ISM at $r=R_{\rm{esc}}=1$ pc, we modify the inner boundary condition to
\begin{align}\label{eq-15}
    \tilde{u}(r)N(r,E)-D(E)\frac{\partial N(r,E)}{\partial r}=\;&-f_{0}\left(\frac{E}{1\,\mathrm{GeV}}\right)^{-\Gamma}\nonumber\\&\times\exp\left(-\frac{E}{E_{c}}\right),
\end{align}
where $f_{0}$ is the normalization in units of GeV$^{-1}$ cm$^{-2}$ s$^{-1}$. The right-hand side of Eq.~(\ref{eq-15}) describes the injection flux of protons into the ISM. Consequently, $f_{0}$ can be derived as
\begin{align}\label{eq-16}
    f_{0}=\;&\frac{\eta_{\rm{esc}}\dot{W}_{p}}{4\pi R_{\rm{esc}}^{2}}\nonumber\\&\times\left[\int_{1\,\mathrm{TeV}}^{\infty}\mathrm{d}E'E'\left(\frac{E'}{1\,\mathrm{GeV}}\right)^{-\Gamma}\exp\left(-\frac{E'}{E_{c}}\right)\right]^{-1}.
\end{align}
Here, $\dot{W}_{p}\approx 3\times10^{38}(R_{W}/Z_{e})$ erg s$^{-1}$ represents the total power of protons above 1 TeV \cite{14}. The term $Z_{e}=1.0\times10^{11}$ cm denotes the height from the emission site to the core \cite{14}. We set the outer extent of proton diffusion in the ISM to $R_{\rm{diff}}=950$ pc, corresponding to $6^{\circ}$ at a distance of 9.0 kpc, and adopt a free-escape condition at the outer boundary $r=R_{\rm{diff}}$.

Using the aforementioned approach, we numerically solve Eq.~(\ref{eq-1}) and obtain the steady-state solution. In the computation, we adopt $\eta_{\rm{esc}}=0.009$ for the $pp+p\gamma_{\rm{UV}}$ scenario and $\eta_{\rm{esc}}=0.005$ for the $p\gamma_{\rm{X}}+p\gamma_{\rm{UV}}$ scenario. The diffusion coefficient in the ISM is assumed to be $D(E)=4.0\times10^{26}(E/1\,\mathrm{GeV})^{1/3}$ cm$^{2}$ s$^{-1}$, identical to that used to simulate GCR diffusion outside the Cygnus OB2 SB. Since Cygnus X-3 is usually regarded as a microquasar, such slow diffusion near a microquasar is supported by a recent study \cite{39}, which indicates that diffusion in the vicinity of a microquasar should be suppressed by one to two orders of magnitude. The ISM density is set to 1 cm$^{-3}$, a typical value for the Galactic plane \cite{40}. After checking the steady-state proton distribution, we find that even protons above 1 PeV can diffuse only to approximately $3^{\circ}$ from the center; therefore, adopting a free-escape condition at $R_{\rm{diff}}$ is reasonable.

The diffuse $\gamma$-ray emission obtained from the steady-state proton spectra is shown in Fig.~\ref{fig:5}. Including this component allows the modeled total $\gamma$-ray spectrum of Cygnus X to explain the multiwavelength observations well. As shown in the figure, the high-energy portion beyond 1 PeV consists of three components: the emission from GCR protons, and the core and diffuse emissions of Cygnus X-3. This result can account for the distribution of $\gamma$-photons above 1 PeV shown in Fig.~1 of \cite{1}. The two photons coming from the direction of Cygnus X-3 can be attributed to its core emission. The other three photons within an angular distance of $\sim3^{\circ}$ from Cygnus X-3 arise from its diffuse emission, whereas the outermost photons originate from GCR protons. 

\section{The accompanying neutrino emission}\label{corr}

\begin{figure}[!t]
    \centering
    \includegraphics[width=0.5\textwidth]{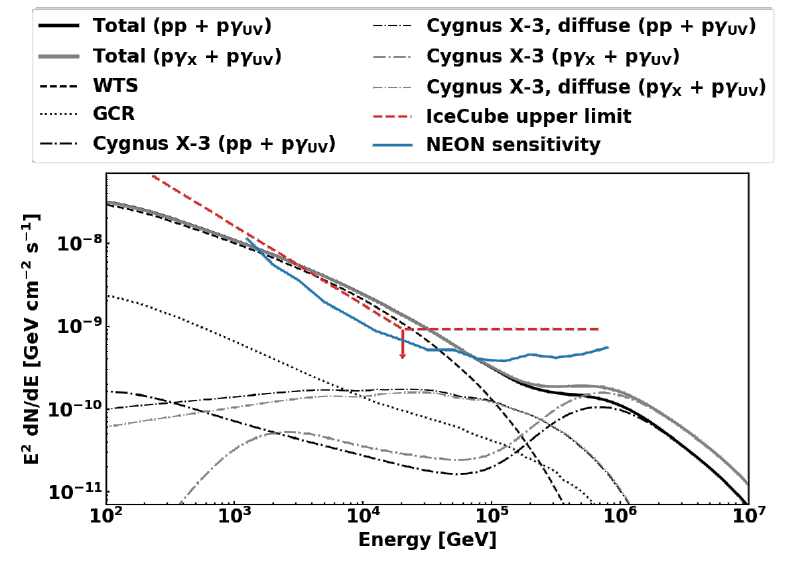}
    \caption{\label{fig:6}Modeled total neutrino emission from Cygnus X, combining contributions from Cygnus OB2, GCR, and Cygnus X-3. The black dashed and dotted lines show the neutrino fluxes from WTS-accelerated protons and GCR protons, respectively. The thick and thin black dash-dotted lines indicate the neutrino fluxes from the Cygnus X-3 core and diffuse emission in the $pp+p\gamma_{\rm{UV}}$ scenario, whereas the corresponding grey lines represent the results for the $p\gamma_{\rm{X}}+p\gamma_{\rm{UV}}$ scenario. The black and grey solid lines give the total neutrino emission from Cygnus X for the two scenarios, respectively. The red dashed line denotes the upper limit on the point source 2HWC J2031+415 from IceCube point source analysis \cite{42}, based on 10 years of IceCube data \cite{51}. The blue solid line is the $3\sigma$ sensitivity of NEON for muon neutrinos with a deployment depth of 3500 m \cite{44}.}
\end{figure}

Since we assume that the $\gamma$-ray emission from the Cygnus X region has a hadronic origin, the prospects for detecting the accompanying neutrino flux with current and next-generation neutrino observatories should also be considered. Given the expected combined origin of the Cygnus X $\gamma$-ray emission\textemdash including Cygnus OB2, GCR, and Cygnus X-3\textemdash we calculate the neutrino emission from each component and sum these contributions to evaluate the total neutrino flux of Cygnus X. In these calculations, all parameters are kept identical to those used in the $\gamma$-ray calculations. For neutrinos produced via $pp$ interactions, we adopt the widely used analytical model of \cite{41}; for $p\gamma$ interactions, we still use the simplified model B of \cite{36}.

The result for a single neutrino flavor is presented in Fig.~\ref{fig:6} and compared with the upper limit on 2HWC J2031+415\textemdash a HAWC-detected point source coincident with the Cygnus OB2 SB \cite{7}\textemdash derived from the publicly available 10-year IceCube dataset \cite{42,51}. This upper limit is scaled down by a factor of 3 to represent the single neutrino flavor scenario. The muon neutrino sensitivity of one of the next-generation neutrino observatories\textemdash the NEutrino Observatory in the Nanhai (NEON) \cite{43,44}\textemdash is also plotted in the figure for comparison. As shown, the expected neutrino flux from our model for both the $pp+p\gamma_{\rm{UV}}$ and $p\gamma_{\rm{X}}+p\gamma_{\rm{UV}}$ scenarios is consistent with the upper limit based on 10 years of IceCube data and significantly exceeds the sensitivity of NEON at 1--100 TeV. With improved techniques, we expect next-generation neutrino observatories to provide a more precise neutrino flux for Cygnus X and to identify whether it has a hadronic origin.

\section{Discussion and Conclusion}\label{dis}
The recent detection of high-energy $\gamma$-ray emission up to PeV energies from the Cygnus X region indicates the existence of a PeVatron therein \cite{1}. The two most promising candidates are Cygnus OB2 and Cygnus X-3. As the nearest and most massive young star cluster in the Milky Way, Cygnus OB2 is a viable site for accelerating particles to PeV energies through multiple channels. Its large population of young massive stars can drive strong stellar winds and produce TSs, which are powerful particle accelerators. Some studies suggest that supernova explosions may also contribute, in which particles are accelerated at the outer blast wave of an SNR \cite{45,46}. The abundant gas in the vicinity of Cygnus OB2 provides a favorable environment for $\gamma$-ray production via $pp$ interactions between high-energy protons and the ambient gas. This makes Cygnus OB2 a plausible source for the $\gamma$-ray emission from Cygnus X. 

In addition to Cygnus OB2, Cygnus X-3 also lies along the line of sight to the Cygnus X region, albeit at a different distance from Earth. Its $\gamma$-rays can therefore also be regarded as contributing to the $\gamma$-ray emission observed from the direction of Cygnus X. Cygnus X-3 is an X-ray binary consisting of a compact object\textemdash likely a black hole or a neutron star\textemdash and a Wolf--Rayet star. Its orbital motion is expected to modulate its multiwavelength emission, and such modulation has been observed in the radio, infrared, X-ray, and GeV $\gamma$-ray bands \cite{10,11,12}. The strong jet kinetic power inferred from X-ray observations, together with the recent LHAASO detection of orbitally modulated PeV $\gamma$-rays from the direction of Cygnus X-3 \cite{14}, provides strong evidence that Cygnus X-3 is also capable of accelerating particles to PeV energies. Its contribution to the Cygnus X $\gamma$-ray emission is thereby non-negligible. 

Recent studies have discussed the prospect that Cyg-nus OB2 or Cygnus X-3 is the PeVatron powering the $\gamma$-ray emission from Cygnus X \cite{16,33,20,32}. In \cite{20} and \cite{32}, the WTS of Cygnus OB2 is adopted as the main accelerator. Under the assumption of suppressed diffusion, high-energy protons accelerated at the TS, together with GCR protons, produce $\gamma$-ray fluxes extending to the PeV band via $pp$ collisions with ambient gas and appear to reproduce the multiwavelength observations, although a deficit remains at PeV energies even allowing for the large error bars. The work \cite{16} claims that Cygnus X-3 is a more natural source above 400 TeV; however, this source alone is insufficient to explain the portion of the spectrum below 200 TeV. These studies suggest that neither Cygnus OB2 nor Cygnus X-3 alone should be regarded as the sole accelerator in Cygnus X.

In this work, we attempt to interpret the $\gamma$-ray emission of the Cygnus X region with a combined scenario that considers Cygnus OB2, Cygnus X-3, and GCR. For Cygnus OB2, we assume that particles are accelerated at the TS and simulate their acceleration and propagation with a conventional transport equation. Under Kolmogorov or Kraichnan diffusion, we find that the steady-state proton cutoff energy is far below the PeV range because of rapid diffusive escape from the Cygnus OB2 SB in these regimes. Even with Bohm diffusion adopted, a significant deficit still remains at energies above 100 TeV. These results suggest that the WTS of Cygnus OB2 is unlikely to be the PeVatron of Cygnus X. For GCR protons, we use the same transport equation to calculate their diffusion near the Cygnus OB2 SB, while the boundary conditions are reconsidered. With a diffusion coefficient much lower than the Galactic value, the GCR contribution can alleviate the shortfall at energies of hundreds of TeV; however, a deficit still remains around PeV, similar to previous studies \cite{20,32}. Therefore, the contribution of Cygnus X-3 appears to be essential. We calculate the $\gamma$-ray emission of Cygnus X-3, including both the core emission dominated by the relativistic jet and the diffuse emission generated by nonthermal particles escaping from the jet. The core emission is modeled under the constraints from the LHAASO observation \cite{14}. The possibility of $\gamma$-rays produced by $pp+p\gamma_{\rm{UV}}$ and $p\gamma_{\rm{X}}+p\gamma_{\rm{UV}}$ channels is considered. The corresponding diffuse emission is then computed for both scenarios, again with a diffusion coefficient significantly lower than the Galactic value. By adding the contribution from Cygnus X-3, the multiwavelength spectrum is well reproduced. All parameters used in the above calculations are reasonable. The gas density is derived from the CfA 1.2 m CO survey and the HI4PI survey, and is consistent with the best-fit value from \cite{32}. The suppressed diffusion around Cygnus OB2 and Cygnus X-3 also has observational and theoretical support \cite{23,39}. This makes our results physically plausible.

Our model naturally explains the $\gamma$-ray emission from the Cygnus X region: the observed flux arises from a superposition of contributions from WTS-accelerated protons in Cygnus OB2, both the core and diffuse emissions of Cygnus X-3, and GCR protons. In this picture, at energies above $\sim200$ TeV, Cygnus X-3 dominates, suggesting that it is the more plausible PeVatron, whereas Cygnus OB2 plays a more important role at lower energies. We expect that next-generation $\gamma$-ray telescopes with improved angular resolution, such as the Cherenkov Telescope Array (CTA) \cite{47} and the Large Array of Cherenkov Telescopes (LACT) \cite{48}, will provide more accurate morphological information about the Cygnus X region in different energy bands and test our scenario. The accompanying neutrino flux is also derived. If the hadronic origin holds, the neutrino flux from Cygnus X should significantly exceed the sensitivity of next-generation neutrino observatories in the TeV band; however, detecting point sources such as Cygnus X-3 requires further improvement. The neutrino spectrum extends into the GeV band, underscoring the need for high-sensitivity GeV neutrino observatories. Moreover, although the $pp+p\gamma_{\rm{UV}}$ and $p\gamma_{\rm{X}}+p\gamma_{\rm{UV}}$ scenarios for the Cygnus X-3 core emission are difficult to distinguish by $\gamma$-ray observations owing to the strong $\gamma\gamma$ absorption near the core, their neutrino spectra are distinct. In particular, for the $pp+p\gamma_{\rm{UV}}$ scenario, GeV neutrino emission at a level of $\sim1\times10^{-10}$ GeV cm$^{-2}$ s$^{-1}$ is anticipated. Testing this prediction would further highlight the need for GeV neutrino observatories, while the sensitivity requirement is very challenging.

\begin{acknowledgements}
This work is supported by the National Natural Science Foundation of China (NSFC) grants 12261141691.
\end{acknowledgements}



\begin{thebibliography}{}
%
%
\bibitem{1}
LHAASO Collaboration, An ultrahigh-energy $\gamma$-ray bubble powered by a super PeVatron, Science Bulletin, 69, 449-457 (2024)

\bibitem{49}
The IceCube Collaboration, High-energy neutrino emission from the Milky Way, ArXiv e-prints [arXiv:2607.25966] (2026)

\bibitem{50}
Abbasi R, Ackermann M, Adams J, et al, Measurement of the Neutrino Spectrum along the Galactic Plane with IceCube, ArXiv e-prints [arXiv:2611.12220] (2026)

\bibitem{42}
Neronov A, Semikoz D, Savchenko D, Neutrino signal from Cygnus region of the Milky Way, Physical Review D, 110, 043024 (2024)

\bibitem{8}
Cao Z, Aharonian FA, An Q, et al, Ultrahigh-energy photons up to 1.4 petaelectronvolts from 12 $\gamma$-ray Galactic sources, Nature, 594, 33-36 (2021)

\bibitem{2}
Wright N, Drew J, Mohr-Smith M, The massive star population of Cygnus OB2, Monthly Notice of the Royal Astronomical Society, 449, 741-760 (2015)

\bibitem{3}
Dame T, Hartmann D, Thaddeus P, The Milky Way in Molecular Clouds: A New Complete CO Survey, The Astrophysical Journal, 547, 792-813 (2001)

\bibitem{4}
HI4PI Collaboration, et al, HI4PI: A full-sky HI survey based on EBHIS and GASS, Astronomy and Astrophysics, 594, A116 (2016)

\bibitem{5}
Ackermann M, Ajello M, Allafort A, et al, A Cocoon of Freshly Accelerated Cosmic Rays Detected by Fermi in the Cygnus Superbubble, Science, 334, 1103-1107 (2011)

\bibitem{6}
Bartoli B, Bernardini P, Bi XJ, et al, Identification of the TeV gamma-ray source ARGO J2031+4157 with the Cygnus Cocoon, The Astrophysical Journal, 790, 152-156 (2014)

\bibitem{7}
Abeysekara AU, Albert A, Alfaro R, et al, HAWC observations of the acceleration of very-high-energy cosmic rays in the Cygnus Cocoon, Nature Astronomy, 5, 465-471 (2021)

\bibitem{9}
Mioduszewski AJ, Rupen MP, Hjellming RM, et al, A Onesided Highly Relativistic Jet from Cygnus X-3, The Astrophysical Journal, 553, 766-775 (2001)

\bibitem{10}
Becklin EE, Neugebauer G, Hawkins FJ, et al, Infrared and X-ray Variability of Cyg X-3, Nature, 245, 302-304 (1973)

\bibitem{11}
Molnar LA, Reid MJ, Grindlay JE, Low-level radio flares from Cygnus X-3, Nature, 310, 662-665 (1984)

\bibitem{12}
Fermi-LAT Collaboration, Abdo AA, Ackermann M, et al, Modulated High-Energy Gamma-Ray Emission from the Microquasar Cygnus X-3, Science, 326, 1512-1516 (2009)

\bibitem{13}
Veledina A, Muleri F, Poutanen J, et al, Cygnus X-3 revealed as a Galactic ultraluminous X-ray source by IXPE, Nature Astronomy, 8, 1031-1046 (2024)

\bibitem{14}
LHAASO Collaboration, Cygnus X-3: a variable petaelectronvolt $\gamma$-ray source, National Science Review, 13, nwag435, (2026)

\bibitem{15}
Wang J, Reville B, Aharonian FA, Galactic Superaccreting X-Ray Binaries as Super-PeVatron Accelerators, The Astrophysical Journal Letters, 989, L25-L32 (2025)

\bibitem{16}
Shi Z, Wang G, Yang RZ, Aharonian FA, Microquasar Cygnus X-3 as the PeVatron Powering the Cygnus Bubble, The Astrophysical Journal Letters, 1006, L31-L38 (2025)

\bibitem{33}
Kachelriess M, Lammert E, Cygnus X-3 as a PeVatron and the LHAASO 2025 data, ArXiv e-prints [arXiv:2512.18786] (2026)

\bibitem{20}
Li B, Blasi P, Amato E, et al, Suppressed diffusion and gamma-ray emission from the Cygnus bubble, Astronomy and Astrophysics, 712, A41 (2026)

\bibitem{17}
Morlino G, Blasi P, Peretti E, et al, Particle acceleration in winds of star clusters, Monthly Notice of the Royal Astronomical Society, 504, 6096-6105 (2021)

\bibitem{18}
Blasi P, Morlino G, High-energy cosmic rays and gamma-rays from star clusters: the case of Cygnus OB2, Monthly Notice of the Royal Astronomical Society, 523, 4015-4028 (2023)

\bibitem{19}
Shi Z, Yang RZ, On the hadronic origin of the very high energy $\gamma$-ray emission surrounding the young massive stellar cluster Westerlund 1, Journal of High Energy Astrophysics, 51, 100560 (2026)

\bibitem{23}
Pandey P, Lopez LA, Rosen AL, et al, Constraining the Diffusion Coefficient and Cosmic-Ray Acceleration Efficiency Using Gamma-Ray Emission from the Star-forming Region RCW 38, The Astrophysical Journal, 976, 98-108 (2024)

\bibitem{55}
Kn\"{o}dlseder J, Cygnus OB2 - a young globular cluster in the Milky Way, Astronomy and Astrophysics, 360, 539-548 (2000)

\bibitem{32}
Roy A, Morlino G, Rishi C, A Multi-messenger Study of the Cygnus bubble, ArXiv e-prints [arXiv:2611.03248] (2026)

\bibitem{21}
Wittor D, Jones T, Vazza F, Evolution of vorticity and enstrophy in the intracluster medium, Monthly Notice of the Royal Astronomical Society, 471, 3212-3225 (2017)

\bibitem{22}
Hahn J, GAMERA - a new modeling package for non-thermal spectral modeling, in 34th International Cosmic Ray Conference (ICRC2015), 34, 917 (2015)

\bibitem{24}
Bisschoff D, Potgieter MS, Aslam OPM, New Very Local Interstellar Spectra for Electrons, Positrons, Protons, and Light Cosmic Ray Nuclei, The Astrophysical Journal, 878, 59-67 (2019)

\bibitem{25}
Cummings AC, Stone EC, Heikkila BC, et al, GALACTIC COSMIC RAYS IN THE LOCAL INTERSTELLAR MEDIUM: \textit{VOYAGER 1} OBSERVATIONS AND MODEL RESULTS, The Astrophysical Journal, 831, 18-38 (2016)

\bibitem{26}
DAMPE Collaboration, Measurement of the cosmic ray proton spectrum from
40 GeV to 100 TeV with the DAMPE satellite, Science Advances, 5, eaax3793 (2019)

\bibitem{27}
Yoon YS, Anderson T, Barrau A, et al, Proton and Helium Spectra from the CREAM-III Flight, The Astrophysical Journal, 839, 5-12 (2017)

\bibitem{28}
LHAASO Collaboration, Precise measurements of the cosmic ray proton energy spectrum in the "knee" region, Science Bulletin, 70, 4173-4180 (2025)

\bibitem{29}
Aartsen MG, Ackermann M, Adams J, et al, Cosmic ray spectrum and composition from PeV to EeV using 3 years of data from IceTop and IceCube, Physical Review D, 100, 082002 (2019)

\bibitem{30}
Shen Z, Yang H, Zuo P, et al, Solar Modulation of Galactic Cosmic-Ray Protons Based on a Modified Force-field Approach, The Astrophysical Journal, 921, 109-115 (2021)

\bibitem{31}
Cao Z, Aharonian FA, An Q, et al, Measurement of Ultra-High-Energy Diffuse Gamma-Ray Emission of the Galactic Plane from 10 TeV to 1 PeV with LHAASO-KM2A, Physical Review Letters, 131, 151001 (2023)

\bibitem{52}
Taylor AR, Gibson SJ, Peracaula M, et al, The Canadian Galactic Plane Survey, The Astrophysical Journal, 125, 3145-3164 (2003)

\bibitem{53}
Bennett CL, Larson D, Weiland JL, et al, NINE-YEAR WILKINSON MICROWAVE ANISOTROPY PROBE (WMAP) OBSERVATIONS: FINAL MAPS AND RESULTS, The Astrophysical Journal Supplement Series, 208, 20-73 (2013)

\bibitem{54}
Planck Collaboration, Planck 2015 results. X. Diffuse component separation: Foreground maps, Astronomy and Astrophysics, 594, A10 (2016)

\bibitem{34}
Matsuoka M, Kawasaki K, Ueno S, et al, The MAXI Mission on the ISS: Science and Instruments for Monitoring All-Sky X-Ray Images, Publications of the Astronomical Society of Japan, 61, 999-1010, (2009)

\bibitem{35}
Beresteteklii VB, Lifshitz EM, Pitaevskii LP, Relativistic Quantum Theory, Pergamon Press, Oxford (1971)

\bibitem{36}
H\"{u}mmer S, R\"{u}ger M, Spanier F, et al, Simplified models for photohadronic interactions in cosmic accelerators, The Astrophysical Journal, 721, 630-652, (2010)

\bibitem{37}
Abe K, Abe S, Abhir J, et al, Long-term study of the gamma-ray emission of Cygnus X-3 with MAGIC and \textit{Fermi}–LAT, Journal of High Energy Astrophysics, 53, 100630 (2026)

\bibitem{38}
Bosch-Ramon V, Aharonian FA, Paredes JM, Electromagnetic radiation initiated by hadronic jets from microquasars in the ISM, Astronomy and Astrophysics, 432, 609-618 (2005)

\bibitem{39}
Fujita Y, Takahashi R, Kawanaka N, Diffusive or Ballistic? Distributions and Spectra of PeV Cosmic Rays around Microquasars, The Astrophysical Journal, 998, 226-230 (2026)

\bibitem{40}
Zhu H, Tian W, Li A, et al, The gas-to-extinction ratio and the gas distribution in the Galaxy, Monthly Notice of the Royal Astronomical Society, 471, 3494-3528 (2017)

\bibitem{41}
Kelner SR, Aharonian FA, Bugayov VV, et al, Energy spectra of gamma rays, electrons, and neutrinos produced at proton-proton interactions in the very high energy regime, Physical Review D, 74, 034018 (2006)

\bibitem{51}
Aartsen MG, Ackermann M, Adams J, et al, Time-Integrated Neutrino Source Searches with 10 Years of IceCube Data, Physical Review Letters, 124, 051103 (2020)

\bibitem{43}
Zhang H, Cui Y, Huang Y, et al, A proposed deep sea Neutrino Observatory in the Nanhai, Astroparticle Physics, 171, 103123 (2025)

\bibitem{44}
Xie C, Yang L, Zhang H, et al, NEON experiment: detectability for extragalactic sources and hardware status, in High Energy Astrophysics in Southern Africa (HEASE 2025), 514, 3 (2025)

\bibitem{45}
Sushch I, Blasi P, Brose R, Supernova remnants in super bubbles acting as cosmic ray
accelerators, Astronomy and Astrophysics, 700, A37 (2025)

\bibitem{46}
H\"{a}rer L, Vieu T, Schulze F, et al, Deciphering the gamma-ray emission in the Cygnus region, Astronomy and Astrophysics, 703, A111 (2025)

\bibitem{47}
Acharya BS, Agudo I, AI Samarai I, et al, Science with the cherenkov telescope array, World Scientific, Singapore (2019)

\bibitem{48}
Zhang S, LACT: Status and future plans, Nuclear Instruments and Methods in Physics Research A, 1086, 171373 (2025)

\bibitem{A-1}
Moskalenko IV, Porter TA, Strong AW, Attenuation of Very High Energy Gamma Rays by the Milky Way Interstellar Radiation Field, The Astrophysical Journal Letters, 640, L155-L158 (2006)

\end{thebibliography}


\appendix

\section{Transport equation solving for Cygnus OB2}\label{appA}
By transforming Eq.~(\ref{eq-1}) into a set of tridiagonal equations, we can numerically solve it using the conventional chasing method. During the solution process, we discretize $r$ and $E$ into meshes of 1000 steps each and obtain the steady-state solution at each $r$ and $E$. For the upstream region, we adopt the vanishing flux boundary condition 
\begin{equation}\label{eq-A1}
    \tilde{u}(r)N(r,E)-D(E)\frac{\partial N(r,E)}{\partial r}=0,
\end{equation}
at the inner boundary $r=R_{\rm{sc}}$, while for the downstream region, the free-escape boundary condition $N(r,E)\allowbreak=0$ is imposed at the outer boundary $r=R_{c}$. At $r=R_{s}$, the boundary condition for both the upstream and downstream regions can be written as \cite{18,20}
\begin{align}\label{eq-A2}
    N_{0}(E)=\;&K\left(\frac{E}{E_{\rm{inj}}}\right)^{-\frac{\mathcal{R}+2}{\mathcal{R}-1}}\exp\left\{\int_{E_{\rm{inj}}}^{E}\frac{\mathrm{d}E'}{E'}\frac{3\mathcal{R}}{1-\mathcal{R}}\right.\nonumber\\&\times\left.\left[\mathcal{D}_{1}(E')-1-\frac{\mathcal{D}_{2}(E')}{\mathcal{R}}\right]\right\},
\end{align}
with
\begin{align}\label{eq-A3}
    K=\;&\frac{3\tilde{u}_{1}}{\tilde{u}_{1}-\tilde{u}_{2}}\frac{\eta_{\rm{inj}}n_{g}}{E_{\rm{inj}}},\\
    \label{eq-A4}\mathcal{R}=\;&\frac{\tilde{u}_{1}}{\tilde{u}_{2}},\\
    \label{eq-A5}\mathcal{D}_{1,2}(E)=\;&\left.\frac{D(E)}{\tilde{u}_{1,2}N_{0}(E)}\frac{\partial N(r,E)}{\partial r}\right|_{r=R_{1,2}}.
\end{align}
Here, $R_{1}$ and $R_{2}$ denote the limiting values of $R_{s}$ approached from the upstream and downstream directions, respectively. 

Since $N_{0}(E)$ and $\mathcal{D}_{1,2}(E)$ depend on each other, determining the boundary condition at the TS requires solving Eq.~(\ref{eq-1}) iteratively. We adopt an initial boundary condition for the first iteration,
\begin{equation}\label{eq-A6}
    N_{0}(E)=K\left(\frac{E}{E_{\rm{inj}}}\right)^{-\frac{\mathcal{R}+2}{\mathcal{R}-1}},
\end{equation}
and then solve the equation. In each subsequent iteration, the boundary condition is updated using the result from the preceding iteration. This procedure is repeated until the solution converges. In our case, convergence typically requires less than 20 iterations.

\section{The $\gamma$-ray emission computing}\label{appB}
The $r$-dependent proton spectra can be obtained by solving Eq.~(\ref{eq-1}). Given these spectra, the corresponding $\gamma$-ray spectra from $pp$ interactions can be computed using the \texttt{GAMERA} toolkit. As discussed in \ref{appA}, we discretize $r$ into a mesh of 1000 steps, which means that the spherical emission site is divided into 1000 spherical shells. The derived $N(r,E)$ can be converted into the total particle number per unit energy contained in the spherical shell between $r-\mathrm{d}r/2$ and $r+\mathrm{d}r/2$ by multiplying it by $4\pi r^{2}\mathrm{d}r$; the result then serves as input to \texttt{GAMERA}. The output of \texttt{GAMERA} is the $\gamma$-ray flux at Earth in units of erg cm$^{-2}$ s$^{-1}$, which represents the total $\gamma$-ray emission from that spherical shell. The average distance $\bar{d}$ to Earth used in the computation is given by 
\begin{align}\label{eq-B7}             \bar{d}&=\sqrt{2\left(\int_{0}^{\pi}\mathrm{d}\theta\frac{\sin\theta}{d_{0}^{2}+r^{2}-2d_{0}r\cos\theta}\right)^{-1}}\nonumber\\&=\sqrt{2d_{0}r\left(\ln\left|\frac{d_{0}+r}{d_{0}-r}\right|\right)^{-1}},
\end{align}
where $d_{0}$ is the distance from the center of the spherical shell to Earth. Summing the $\gamma$-ray flux from each spherical shell thus yields the total $\gamma$-ray emission of Cygnus X.

The $\gamma$-rays undergo $\gamma\gamma$ absorption by the interstellar radiation field (ISRF) and Cosmic Microwave Background (CMB) during their propagation through the interstellar medium to Earth. For Cygnus X, the contribution of the ISRF to the absorption optical depth is negligible compared with that of the CMB \cite{14}. Thus, the optical depth for this interstellar absorption can be approximated as \cite{A-1}
\begin{align}\label{eq-B8}
    \tau_{\gamma\gamma}(E)=\;&-\frac{32\pi dk_{B}T_{\rm{CMB}}}{h^{3}c^{3}E^{2}}\int_{m_{e}c^{2}}^{\infty}\mathrm{d}E_{t}E_{t}^{3}\sigma_{\gamma\gamma}(E_{t},E_{t})\nonumber\\&\times\ln\left[1-\exp\left(\frac{E_{t}^{2}}{Ek_{B}T_{\rm{CMB}}}\right)\right].
\end{align}
Here, $d$ is the distance from the emission site to Earth and $T_{\rm{CMB}}$ is the CMB temperature. The absorption cross section $\sigma_{\gamma\gamma}(E_{t},E_{t})$ has the same form as in Eq.~(\ref{eq-13}), with the first argument $E$ replaced by $E_{t}$.

\end{document}